\documentclass[conference]{IEEEtran}
\IEEEoverridecommandlockouts
\usepackage{cite}
\usepackage{amsmath,amssymb,amsfonts}
\usepackage{algorithmic}
\usepackage{graphicx}
\usepackage{textcomp}
\usepackage{xcolor}
\def\BibTeX{{\rm B\kern-.05em{\sc i\kern-.025em b}\kern-.08em
    T\kern-.1667em\lower.7ex\hbox{E}\kern-.125emX}}
\begin{document}

\title{Requirements' Characteristics: How do they Impact on Project Budget in a Systems Engineering Context?\\
}

\author{\IEEEauthorblockN{Panagiota Chatzipetrou}
\IEEEauthorblockA{\textit{Department of Informatics, CERIS} \\
\textit{ \"Orebro University School of Business,}\\
SE-701 82, \"Orebro, Sweden \\
panagiota.chatzipetrou@oru.se}
\and
\IEEEauthorblockN{Michael Unterkalmsteiner}
\IEEEauthorblockA{\textit{SERL - Sweden} \\
\textit{Blekinge Institute of Technology}\\
SE-371 41, Karlskrona, Sweden \\
michael.unterkalmsteiner@bth.se}
\and
\IEEEauthorblockN{Tony Gorschek}
\IEEEauthorblockA{\textit{SERL - Sweden} \\
\textit{Blekinge Institute of Technology}\\
SE-371 41, Karlskrona, Sweden \\
tony.gorschek@bth.se}

}

\maketitle

\begin{abstract}
\textbf{Background}: Requirements engineering is of a principal importance when starting a new project. However, the number of the requirements involved in a single project can reach up to thousands. Controlling and assuring the quality of natural language requirements (NLRs), in these quantities, is challenging.  

\textbf{Aims}: In a field study, we investigated with the Swedish Transportation Agency (STA) to what extent the characteristics of requirements had an influence on change requests and budget changes in the project. 

\textbf{Method}: We choose the following models to characterize system requirements formulated in natural language: Concern-based Model of Requirements (CMR), Requirements Abstractions Model (RAM) and Software-Hardware model (SHM). The classification of the NLRs was conducted by the three authors. The robust statistical measure Fleiss' Kappa was used to verify the reliability of the results. We used descriptive statistics, contingency tables, results from the Chi-Square test of association along with post hoc tests. Finally, a multivariate statistical technique, Correspondence analysis was used in order to provide a means of displaying a set of requirements in two-dimensional graphical form.

\textbf{Results}: The results showed that software requirements are associated with less budget cost than hardware requirements. Moreover, software requirements tend to stay open for a longer period indicating that they are "harder" to handle. Finally, the more discussion or interaction on a change request can lower the actual estimated change request cost. 

\textbf{Conclusions}: The results lead us to a need to further investigate the reasons why the software requirements are treated differently from the hardware requirements, interview the project managers, understand better the way those requirements are formulated and propose effective ways of Software management.

\end{abstract}


\section{Introduction}
Requirements specifications written in natural language text are ubiquitous in the public sector where projects are implemented following a request for tender process. Formal contracts between the ordering party and suppliers serve as the basis for formulating requirements statements, which, in turn, are the starting point for more refined technical requirements and design specifications. In other words, natural language requirements bridge the gap between the needs expressed by the public (government) and the solutions that fulfill those needs, designed and implemented by contractors~\cite{ref_unterkalmsteiner2018}.

We have performed a field study at the Swedish Transport Agency (STA) to investigate the characteristics of natural language requirements (NLRs) and their impact on project execution, accounted by change requests on these initially defined NLRs. Our goal was to identify any associations between particular types of requirements and budget changes. This information would be beneficial for STA as it could be used to focus on problematic requirements types, early on in the project, by monitoring their design and implementation more closely. In the studied case, there were dozens of contractors and the requirements were used not only to prepare for tendering, but also to coordinate between all project parties.

In the scientific literature, several different models for analyzing and controlling NLRs have been introduced. We adopted the following three models: the Concern-based Model of Requirements (CMR)~\cite{ref_glinz2007} proposed by Glinz, the Requirements Abstractions Model (RAM)~\cite{ref_gorschek2006} proposed by Gorschek and Wohlin, and the Software-Hardware Model (SHW), which we incorporated from the systems and software engineering standard for architecture descriptions, ISO/IEC/IEEE 42010~\cite{ref_iso2011}. We chose these models because of their general applicability in the systems engineering context and their successful use in the past to characterize requirements. The classification of the NLRs was conducted by the three authors. Each one of the authors independently categorized an equal number of requirements. The robust statistical measure Fleiss' Kappa was used to verify the reliability of the results.  Moreover, we chose to analyse requirements where we had a record on how they developed over time in relation to their budget and the employees' interaction (i.e. written comments on the change requests). 

In our study we aimed to investigate how change requests to NLRs impact a project's budget changes, but also to understand and how system requirements in large infrastructure projects are managed over time. For that reason, we used an exploratory case study, where we applied and presented results from descriptive statistics, contingency tables, Chi-Square test of association along with post hoc tests. In addition, a multivariate statistical technique, Correspondence analysis was used, in order to illustrate the sets of requirements in two-dimensional graphical form. The results showed that Software requirements are treated differently from the Hardware requirements, in particular they are associated with less budget while they are more difficult to handle.

The remainder of the paper is structured as follows: Section~\ref{sec:rel} provides an outline of the related work. Section \ref{sec:models} provides a brief description of the requirements' characterization models that were used in the study. Section~\ref{sec:rm} presents the research methodology of our work. Section~\ref{sec:res} presents the results of the analysis. Finally, in Section~\ref{sec:con} discussion and conclusions are provided.

\section{Related Work}\label{sec:rel}
The association between requirements engineering and other activities in the software product development process, project success, and product quality has been the focus of several studies in the past. Damian and Chisan~\cite{ref_damian2006} investigated in a case study how improvements in the requirements engineering process affect other development processes. They have observed that the introduced practices (feature decomposition, requirements traceability, group analysis sessions, cross functional teams, structured requirements, testing according to requirements) had payoffs in increased productivity, quality and improved risk management. Zowghi and Nurmuliani~\cite{ref_zowghi2002} conducted a survey among 52 software development companies in Australia, studying the relationship between requirements volatility, i.e. the potential for change in the business environment or the stakeholders' fluctuation in the understanding of the requirements, and project performance measured by accuracy of schedule and cost estimates. They found support for their hypothesis that requirements' volatility leads to budget and schedule overruns.

Particular errors in requirements specifications and their impact on project outcomes have been also studied. Veras et al.~\cite{ref_veras2010} classified 2188 requirements from the aerospace domain and found a surprisingly high number of defective requirements (10\%), the most frequent types of errors being external conflicts/consistency, traceability, external completeness and requirement completeness. While the authors suggested application scenarios for their findings and proposed classification (training requirements reviewers, requirements error estimation, checklists for requirements verification, benchmarking requirements specifications), they did not investigate the impact of requirements errors and their type on downstream development. Chari and Agrawal~\cite{ref_chari2018} studied how change requests on incorrect or incomplete requirements affect software quality and development effort. They analyzed data from 49 management information system projects and found that, while the resolution of incorrect requirements led to fewer defects, new requirements were generated at the same time which were associated with an increase in delivered defects and effort. Similarly, Kamata and Tamai~\cite{ref_kamata2007} investigated in 72 projects of a software company in Japan to what degree requirements quality is associated with project performance. They found that the quality of the introductory section of a requirements specification (purpose, scope, definitions, references, overview) is associated with project performance in terms of cost overruns. Unfortunately, the authors did not illustrate how the organizations quality assurance team assessed the quality of requirements.

All these studies provide valuable insight on how particular requirements engineering aspects (requirements management activities, requirements quality, and requirements errors), affect downstream development. The purpose of the study presented in this paper is to extend that body of knowledge with the perspective of requirement types, classified according to the models presented in Section~\ref{sec:models}. Moreover, we investigate the impact of the requirements’ inherent characteristics, i.e. the amount of frequency of discussions and overall analysis time, on budget changes. Therefore, the contribution of the paper is twofold: first to describe the process and the design of the study. The main contribution is to understand and reason the management of the Software requirements. The case study was applied in a real industry data so as to draw interesting and usefully conclusions regarding the requirements management in relation with the project budget.

\section{Requirements characterization models}\label{sec:models}
The purpose of classifying a set of requirements according to a given model is to enhance our understanding of the nature of the requirements that are specified in large-scale, infrastructure projects. We therefore chose models that were simple enough (structurally and conceptually) to be applied to realistic requirement statements that were written by stakeholders unaware of to these models. Another criterion was that the chosen models contain operative guidelines or rules on how to classify requirements.

\subsection{Concern-based Model of Requirements (CMR)}
Glinz~\cite{ref_glinz2007} proposes classifying requirements into four categories: 
\begin{enumerate}
\item \textit{Functional requirements}, which are related to functional concern
\item \textit{Non Functional Requirements}
\begin{itemize}
\item \textit{Performance requirements} which are related to a performance concern,
\item \textit{Specific quality requirements} which are related to a quality concern 
\end{itemize}
\item \textit{Constraints} which refer to requirements that constrains the
solution space beyond what is necessary for meeting the given functional, performance, and specific quality requirements.
\end{enumerate}

Glinz~\cite{ref_glinz2007} suggests using the following questions, applied in this order, to classify requirements: 

Does the requirement specify...
\begin{enumerate}
    \item ... some of the system's behavior, data, input, or reaction to input stimuli - regardless of the way how this is done? $\implies$ Functional
    \item ... restrictions about timing, processing or reaction speed, data volume, or throughput? $\implies$ Performance
    \item ... a specific quality that the system or component shall have? $\implies$ Specific quality
    \item ... any other restriction about what the system shall do, how it shall do it, or any prescribed solution or solution element? $\implies$ Constraint
\end{enumerate}

We used these questions, allowing however for requirements to be classified into multiple categories as we soon realized that real-world requirements may cover multiple concerns.\ 

A requirement could be part of more than one category. Thus, when the three authors independently categorized the 215 requirements, they assigned with them 1 or 0 depending on whether the requirement belonged to at the respective category i.e. Functional requirements, Non-Functional or Quality requirements, Constrain requirements, Functional AND Quality requirements e.t.c. 

\subsection{Requirements Abstraction Model (RAM)}
Gorschek and Wohlin~\cite{ref_gorschek2006} propose classifying requirements in four abstraction levels: product, feature,  function and component level. 

\begin{enumerate}
\item \textit{Product Level}: requirements on this, most abstract, level do not fitting the normal definition of a requirement (e.g., testable and unambiguous). Product Level requirements are considered abstract enough to be comparable directly to the product strategies and indirectly to the organizational strategies.
\item \textit{Feature Level}: requirements on this level are features that the product supports, i.e. the requirements are usually an abstract description of the feature itself.
\item \textit{Function Level}: requirements on the this level describe what a user should be able to do: actions that are possible to perform or non-functional requirements that should be fulfilled.  
\item \textit{Component Level}: requirements on this level are of a detailed nature depicting information that is closer to (or even examples of) how something should be solved.
\end{enumerate}

Similar to the CMR, Gorschek and Wohlin~\cite{ref_gorschek2006} provide a set of questions as a mean to operationalize the model and classify requirements:
\begin{enumerate}
    \item Is the requirement functional or does it describe testable characteristics that the product should have? $\implies$ Functional level
    \item Does the requirement consist of a specific suggestion of \textit{how} something should be solved? $\implies$ Component level
    \item Is the requirement abstract enough to be comparable to the product strategies? $\implies$ Product level
    \item Does the requirements describe a feature that should be supported? $\implies$ Feature level
\end{enumerate}

We used these questions to classify requirements into \textit{Problem oriented} requirement formulations (Product, Feature and Function level) and \textit{Solution oriented} requirement formulations (Component level).

\subsection{Software-Hardware Model (SHM)}
Since we were studying a case from the system engineering context, we were interested in what degree the requirements describe hardware, software and mixed product aspects. Hence, we chose to classify the requirements, using definitions from the systems and software engineering standard, ISO/IEC/IEEE 42010~\cite{ref_iso2011}:
\begin{enumerate}
\item \textit{Software intensive requirements} belong to "any system where software contributes essential influences to the design, construction, deployment, and evolution of the system as a whole". Moreover, requirements that refer to hardware controlled by software are also included in this category.
\item \textit{Hardware intensive requirements} include only hardware actions or properties that are not controlled by software.
\end{enumerate} \ 
\

\section{Research Methodology}\label{sec:rm}
Our work is driven by the following research questions:\
\bigskip

\textit{RQ1: Which requirement characteristics are associated with budget changes within the studied project?}\
\bigskip

We wanted to explore whether particular requirements' characteristics, defined by the models described in Section~\ref{sec:models}, can be associated with budget changes during the project execution.\ 
\bigskip

\textit{RQ2: Which requirements' inherent characteristics are associated with budget changes within the studied project?}
\bigskip

We wanted to explore whether particular requirements' inherent characteristics, presented in Section~\ref{sec:dataset}, can be associated with budget changes during the project execution. 
\bigskip

The above RQs are the starting point in our investigation and will help us drive our research and gain further understanding towards whether the requirements' characteristics (inherent or not) affect on the project budget and in which ways.

\subsection{Design of the study}
To guarantee that the requirements will be classified objectively across the three requirements' characterization models, we involved all the three authors into the classification process. To access the reliability of the agreement between the three authors we applied the robust statistical measure Fleiss' Kappa~\cite{ref_kappa,ref_fleiss}. Different classifications have been suggested for assessing how good the strength of agreement is when based on the value of Cohen's kappa coefficient. According to the guidelines from~\cite{ref_Altman} and~\cite{ref_LandisKoch} (Table~\ref{TableKappa}) we considered an agreement when the kappa value is above 0.5.

\begin{figure}[ht]
	\centering
    \includegraphics[width= 1\linewidth]{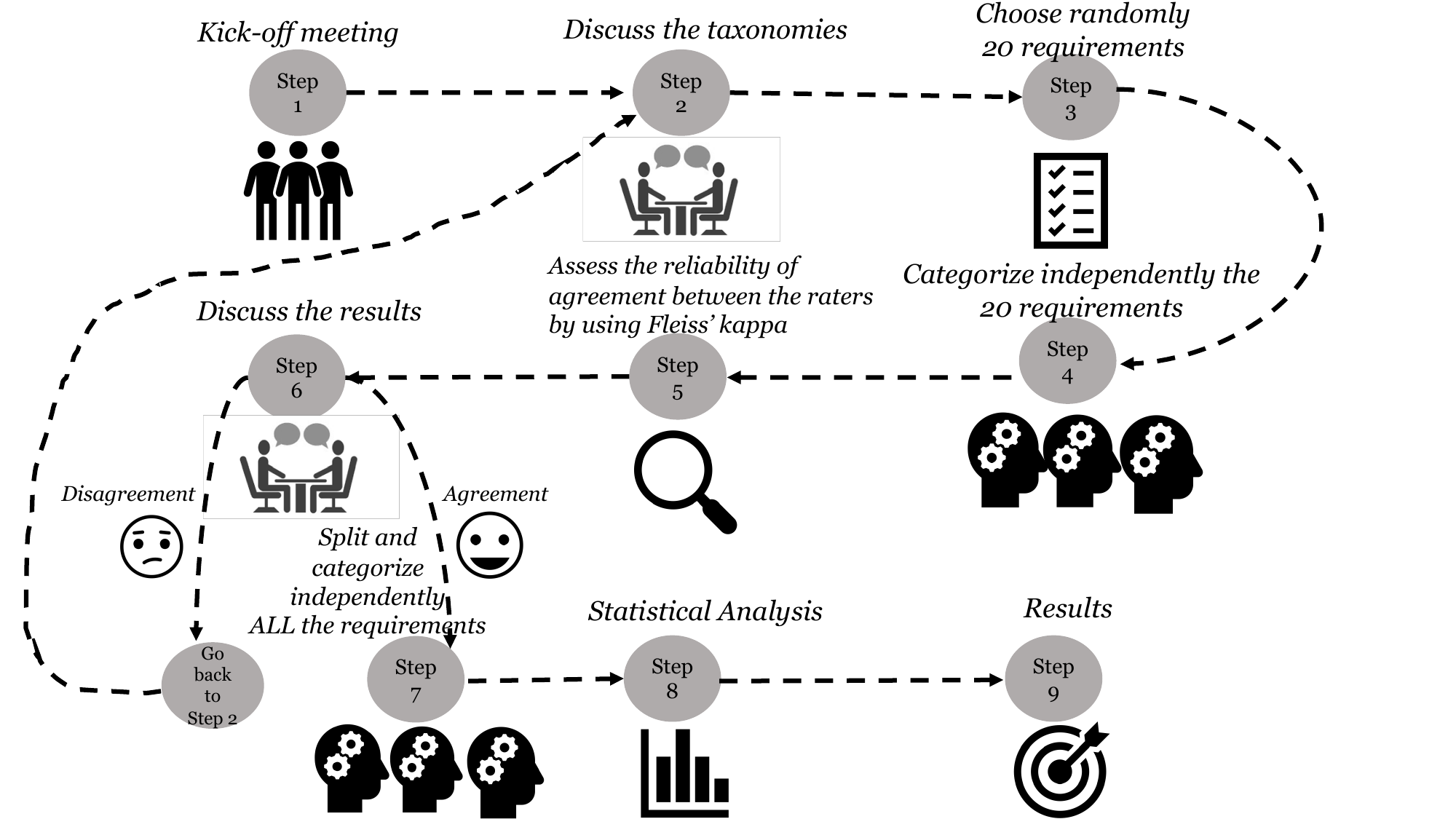}
    \caption{Design of the study}
    \label{fig:Designofthestudy}
\end{figure}

The design of the study is depicted on Fig.~\ref{fig:Designofthestudy}. The three authors had a kick-off meeting where they discussed the nature of the requirements received from the Swedish Transport Agency (STA). At the same meeting they discussed the structure of each one of the three requirements' characterization models in order to assure that all three shared the same understanding. 

At the next step, 20 requirements from our data set were chosen randomly and were independently categorized by the authors. The authors returned one week later with their classifications. The robust statistical measure Fleiss' Kappa~\cite{ref_kappa,ref_fleiss} was applied to access the reliability of the agreement between the three authors. Since we considered an agreement when the kappa value is above 0.5, in the opposite case (Kappa lower than 0.5), we agreed that we disagreed on our classifications and we discussed the requirements' characterization models again. At the end of the meeting a new set of 20 different requirements was again chosen. The three authors repeated the classification, again independently. We repeated the process until we reached to a good agreement (Fleiss' Kappa above 0.5). We reached agreement after 3 iterations. At that point we split the whole data set into three equal parts and each author was assigned with one part. The classification of all the requirements was completed. The statistical analysis and the results are presented in the next section. 

\begin{table}
\centering
\caption{Interpretation of Cohen's Kappa statistic}\label{TableKappa}
\begin{tabular}{|l|l|}
\hline
\textbf{Cohen's Kappa value} & \textbf{Interpretation} \\
\hline
0-0.2 & Poor\\
0.21-0.4 & Fair\\
0.41-0.6 & Moderate\\
0.61-0.8 & Good\\
0.81-1.00 & Very Good\\
\hline
\end{tabular}
\end{table}

\subsection{Data set}\label{sec:dataset}
The studied requirements originate from a large infrastructure project that commenced in 2007 and was finally completed in 2017. 
\vspace{2mm}

\subsubsection{Description of the data set}

 The data set contained important information about the requirements, i.e. budget, comments from the employees, time sequence of change requests and dates on which a change request was decided and the last date on which the requirement was changed. We refer to them as \textit{inherent characteristics} of the requirements. 
 
 Initially our data set contained 5073 requirements. Since our aim was to study the requirements with assigned budget and change requests in relation to the different requirements' models, the original data set was screened to ensure it contained the necessary data for our research.  Specifically: 
 
\begin{itemize}
  \item Requirements with no change requests were excluded,
  \item For investigating the reasons behind the change requests and the amount of budget assigned, we included in our analysis only the requirements which contain comments.
\end{itemize}

 The final data set included 215 requirements, which were used in the subsequent analyses. \ 
 \vspace{2mm}
 
 \subsubsection{Descriptive statistics}\
 
\begin{figure}[ht]
	\centering
	\includegraphics[width= 1\linewidth]{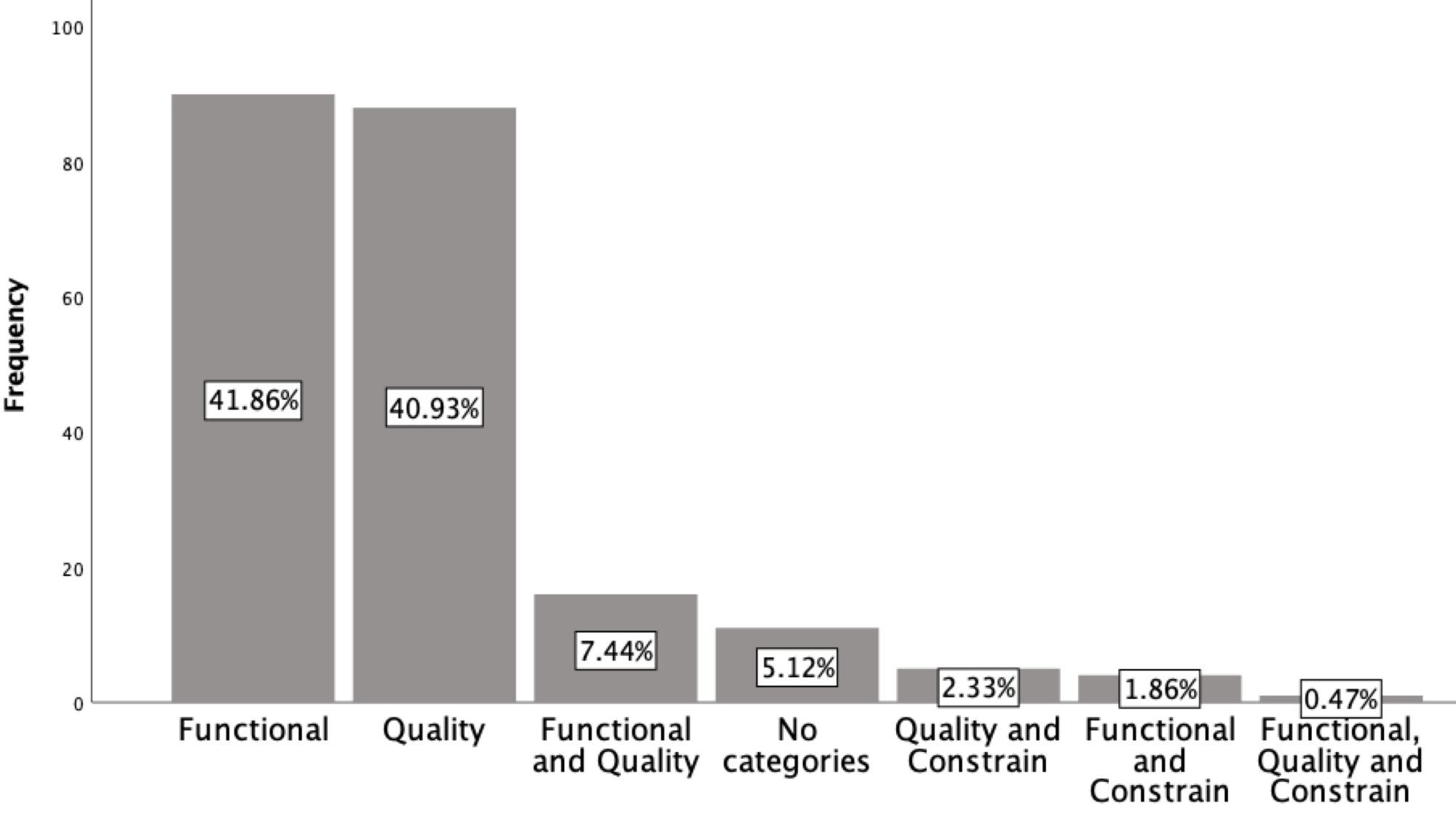}
    \caption{Concern-based Model of Requirements (CMR)}
    \label{fig:Figure_2}
\end{figure}

\begin{figure}[ht]
	\centering
	\includegraphics[width= 1\linewidth]{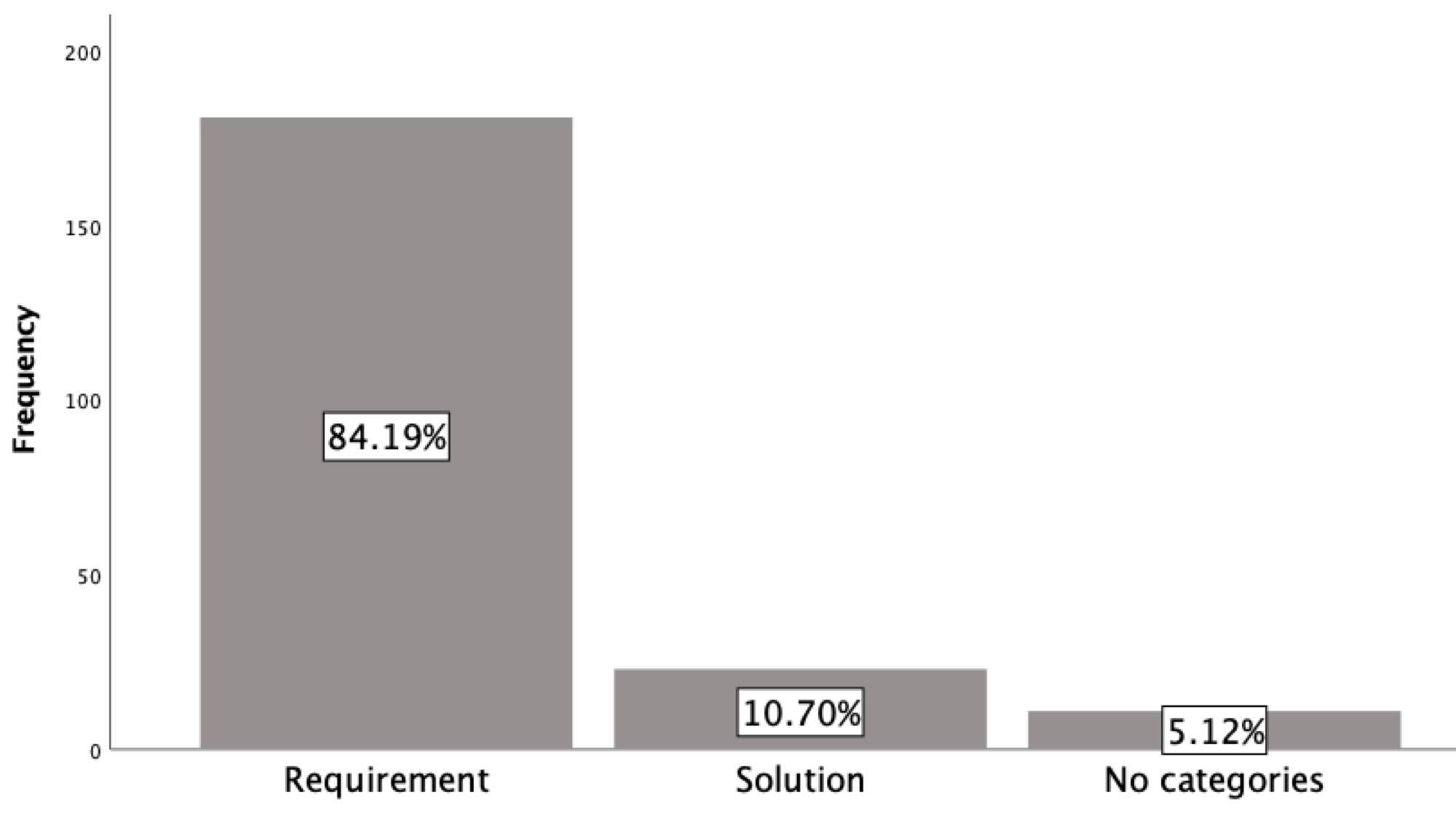}
    \caption{Requirements Abstraction Model (RAM)}
    \label{fig:Figure_3}
\end{figure}

\begin{figure}[ht]
	\centering
	\includegraphics[width= 1\linewidth]{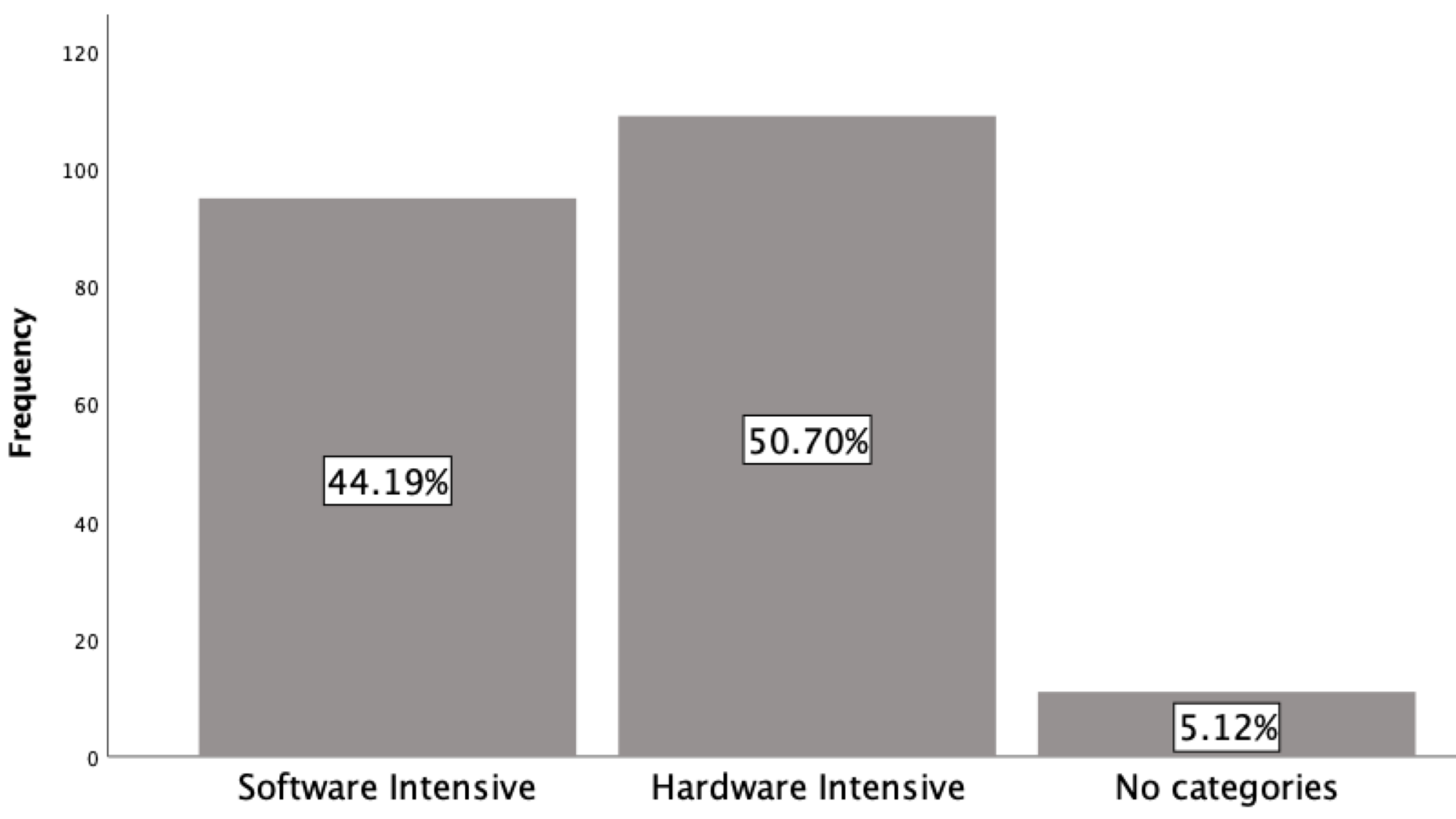}
    \caption{Software intensive - Hardware intensive Model}
    \label{fig:Figure_4}
\end{figure}

 The distribution of the requirements within the three different above mentioned models are available in Fig.~\ref{fig:Figure_2}, Fig.~\ref{fig:Figure_3} and Fig.~\ref{fig:Figure_4}. In particular, regarding the Concern-based Model of Requirements (Fig.~\ref{fig:Figure_2}), the requirements were categorized with almost the same distribution between the two dominant categories, i.e. Functional requirements and Quality requirements, (almost 42\% to each category). The same is true for the Software - Hardware intensive model (Fig.~\ref{fig:Figure_4}). Almost half of the requirements (52\%) are categorized as Hardware, where 44\% is categorized as Software requirements. Among the 4,6\% of the requirements that were not categorized, the most frequent cases included requirements that were vague, irrelevant or implying to belong to both categories. On the other hand, when we categorize the 215 requirements towards the RAM Model (Fig.~\ref{fig:Figure_3}), the majority of the requirements (almost 85\%) were characterized as Requirements, i.e. belong at the Problem level, where only the 10\% was characterized as Solutions, i.e. belong at the last level of abstraction.\

\begin{figure}[ht]
	\centering
	\includegraphics[width= 1\linewidth]{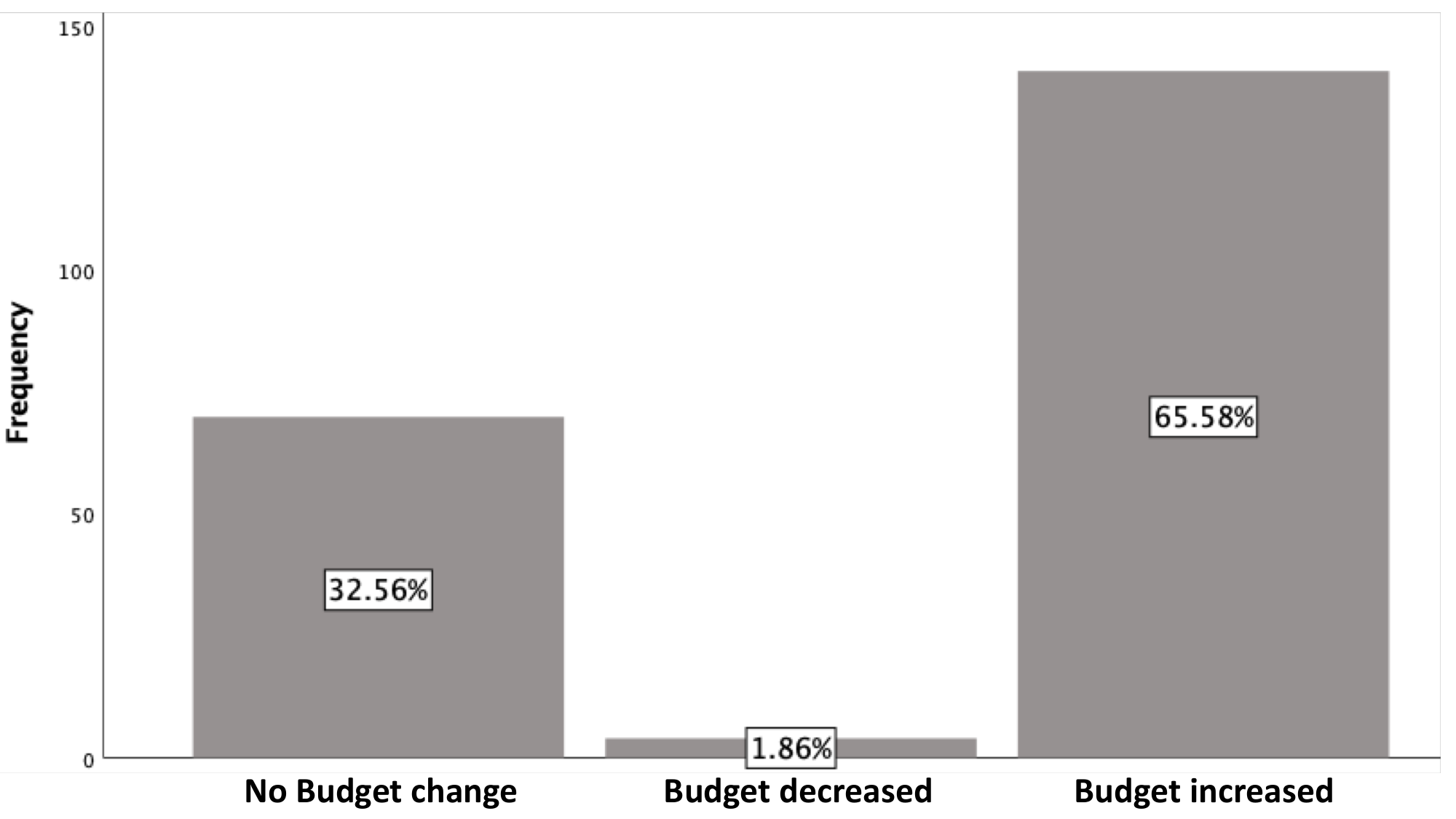}
    \caption{Budget of the requirements with Change requests}
    \label{fig:FigureBudget}
\end{figure}

\begin{figure}[ht]
	\centering
	\includegraphics[width= 1\linewidth]{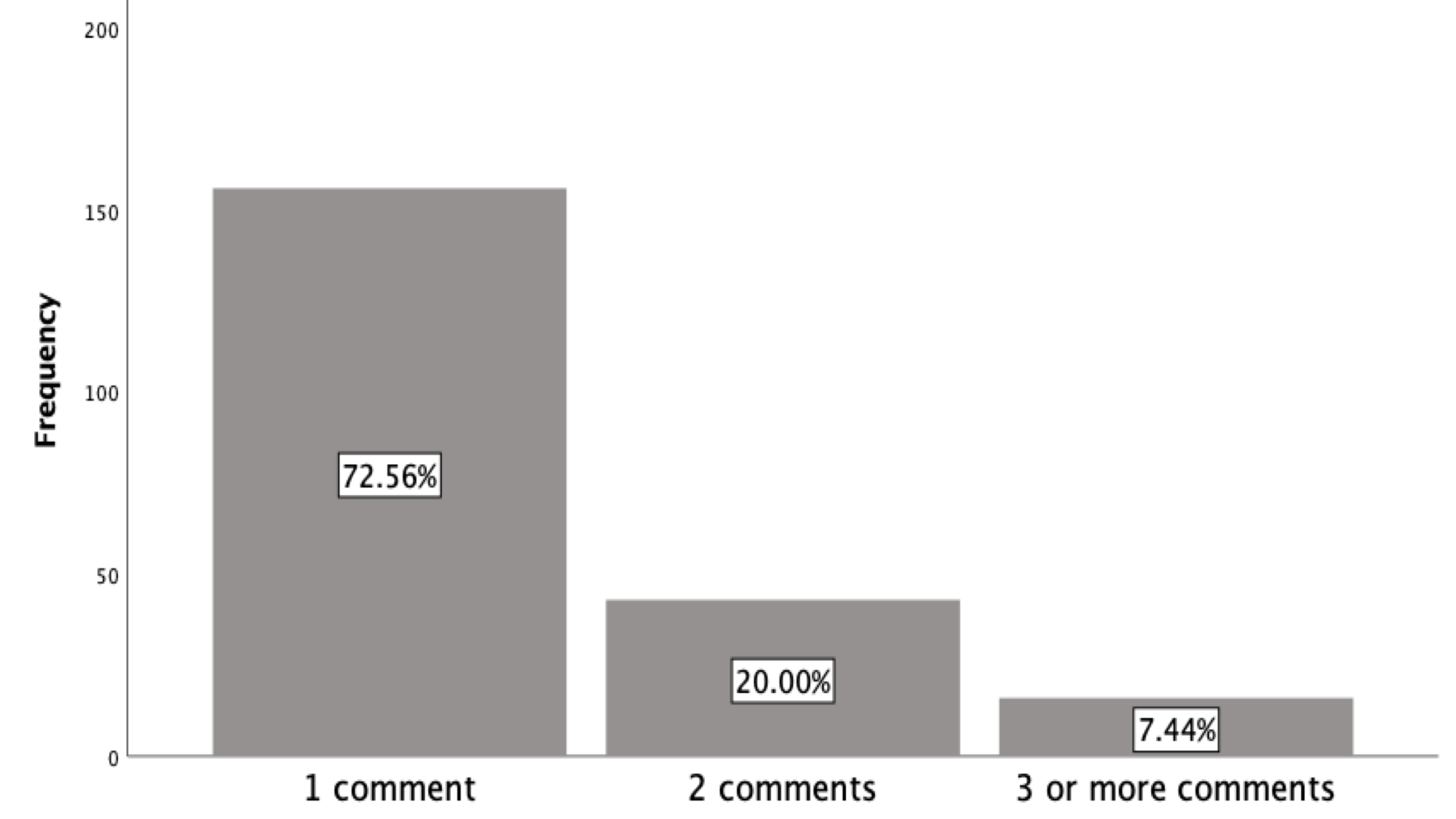}
    \caption{Comments of the employees (in numbers)}
    \label{fig:FigureComments}
\end{figure}

 Furthermore, the important inherent characteristics of the requirements that we used in our analysis are: a) Budget and b) Comments from the employees. The distribution of the requirements within Budget and Comments from the Employees are available in Fig.~\ref{fig:FigureBudget} and Fig.~\ref{fig:FigureComments} respectively. 
 
 The results showed that the majority of the change requests (65,5\%) required the budget to be increased in order to be addressed, while only the 35,5\% of the change requests could be addressed without extra budget cost. Although, we noticed that a small percentage of the change requests (only 2\%) when they were addressed, the project budget was decreased. 
 
 Moreover, we categorize the requirements according to the number of comments the received from the employees. The majority of the requirements received only one comment (72,5\%), where the 20\% received two comments and just 7,4\% received three or more comments. \
 
\begin{figure}[ht]
	\centering
	\includegraphics[width= 1\linewidth]{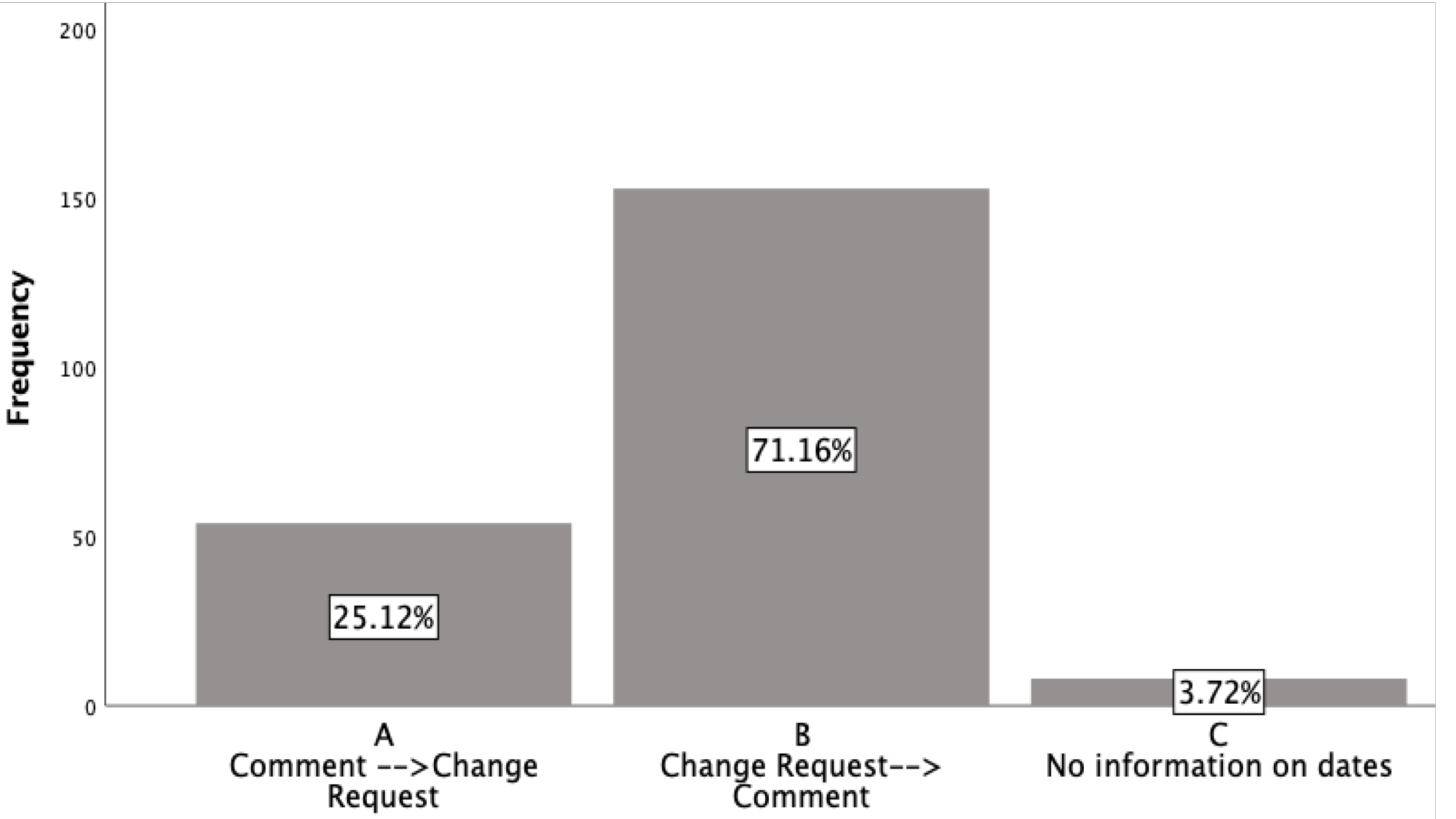}
    \caption{Time sequence between comments and change requests}
    \label{fig:FigureTimeSequence}
\end{figure}

 Furthermore, the extrapolated data were used to calculate two important variables for our study: a) \textit{Analysis Time} and b) \textit{Time sequence of change requests and comments}. \
 
 The (a) \textit{Analysis Time} is the time between the date on which a change request was decided and the last date on which the requirement has changed. Due to the fact that a change in a requirement may have occurred before the change request was decided, negative values exist. The time is computed in total days.\ 
 
To investigate whether employees' comments are crucial and affect the existence of the changes requests we calculated the second variable (b) \textit{Time sequence of change requests and comments} which deals with the chronological order of the date on which the decision of a change request was made and the date on which the first comment was made. 

Three categories were identified (Fig.~\ref{fig:FigureTimeSequence}): 
\begin{itemize}
    \item In case A, first we have one or more comments and then a change request is decided.
    \item In case B, first we have a change request decided and then one, more comments follow.
    \item In case C, we don’t have any information on the dates
\end{itemize}

  The results showed that only the 25\% of the requirements were commented on by employees before a change request was received. Most of the cases were commented on after a change request has occurred (more than 70\%).

\subsection{Data Analysis}

\subsubsection{Chi-square test of association}
To test if there are any associations between the different models' categories and the other factors i.e. Budget, Time, we used the Chi-Square test of association. To prevent Type I errors, we used exact tests, and more specifically, the Monte-Carlo test of statistical significance based on 10 000 sampled tables and assuming (p = 0.05)~\cite{ref_hope1968}. To examine the strength of associations we use Cramer's V test. Cramer's V is a measure of the strength of association of a nominal by nominal relationship. Cramer's V ranges in value from 0 to +1 with a value of 0 indicating no association to a value of 1 indicating complete association. Cohen~\cite{ref_cohen1988} suggested the following guidelines for interpreting Cramer's V (See Table~\ref{TableCramersV}).

However, finding an association did not provide us with further details about this association (e.g., which cases are 'responsible' for this association). Therefore, following up our statistical significant results, we performed post hoc testing using adjusted standardized residuals~\cite{ref_agresti2003, ref_siegel988}. By analyzing these values, we had a cell-by-cell comparison of the expected versus observed frequencies, which helped us to understand which cases where deviated from the independence. We consider an adjusted residual significant if the absolute value is above 1.96, as suggested by~\cite{ref_agresti2003}.\

\vspace{1mm}

\subsubsection{Correspondence Analysis}
Correspondence analysis provides an interpretation approach for illustrating our results. As the name of the method suggests, it is a way to explore the ‘‘system of associations’’ between the elements of two sets. Correspondence Analysis is a statistical visualisation method for picturing the associations between the categorical variables of a two-way contingency table. These relationships are described by projecting the values of the variables as points on a two-dimensional space, in such a way that the resulting plot describes simultaneously the relationships between the variables. For each variable, the distances between points in the plot reflect the relationships between them \cite{ref_correspanalysis}.

\vspace{1mm}

\subsubsection{Non-parametric tests}
Non-parametric tests i.e. Kruskal-Wallis H test was performed to determine if there are statistically significant differences between two or more variables which are on a continuous and categorical level \cite{ref_Kruskal}.

\begin{table}
\centering
\caption{Interpretation of Cramer's V test}\label{TableCramersV}
\begin{tabular}{|l|l|}
\hline
Cramer's V value & Interpretation \\
\hline
0.1 & Weak association\\
0.3 & Moderate association\\
0.5 & Strong association\\
\hline
\end{tabular}
\end{table}

\subsection{Validity Threats}
The validity threats are distinguished between four aspects of validity according to \cite{ref_runeson_host_2009}:

\subsubsection{Construct validity}
Construct validity reflects the extent to which the operational measures represent the study subject. In the present study, all the data were acquired from company archives, thus representing objective measures. No subjective measures were used, such as the ones elicited through interviews or surveys.
\subsubsection{Internal validity}
Internal validity refers to the examination of causal relations, which is the intended outcome of our investigation. In our case study, we investigated on the impact between a number of factors i.e. budget, time and discussion in order to understand how those requirements' characteristics impact on the project budget. 
\subsubsection{External validity}
External validity threat concerns to what extent the results could be valid to other companies. In our case, the study is clearly exploratory and by no means can the findings be generalized to an isolated company. 
\subsubsection{Reliability}
Regarding reliability, this aspect is concerned with to what extent the data and the analysis are dependent on the specific researchers, i.e. avoid respondent's and researcher's personal biases.\

To address this threat, we conducted a number of workshops between the authors. The purpose of those workshops was twofold: First, to ensure that all the authors share the same knowledge and understanding towards the chosen requirements models. Second, to evaluate the reliability of their agreement by assigning them to categorize independently random sets of requirements and assess the reliability of their agreement by using the robust statistical measure Fleiss' Kappa. 
\section{Results}\label{sec:res}

As already mentioned, the study is exploratory, thus, a number of different factors were analyzed in relation with the budget of the change requests. The most interesting results are presented in the present section. \

\subsection{Software - Hardware intensive requirements VS Budget }
\begin{figure}[ht]
	\centering
	\includegraphics[width= 1\linewidth]{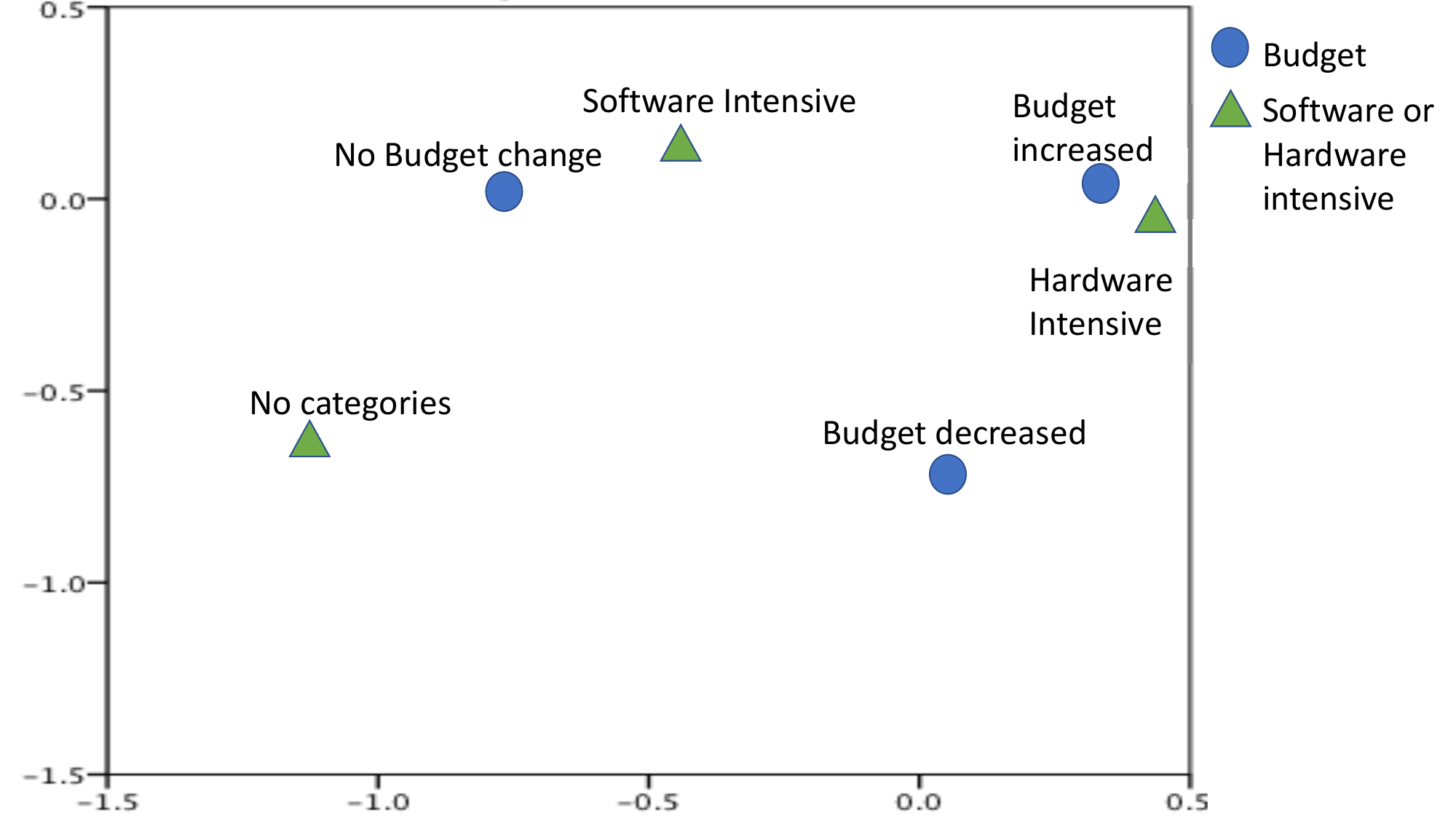}
    \caption{Software - Hardware intensive requirements VS Budget}
    \label{fig:Figure_5}
\end{figure}

The results from the contingency tables and Correspondence analysis (Fig.\ref{fig:Figure_5}) showed that Functional software requirements are associated with less budget change cost than functional hardware requirements ($p<0.05$, Cramer's $V=0.300$). In particular, among the functional requirements, more than 85\% of the hardware intensive requirements were assigned with an increasing budget change while more than half of the software requirements were assigned with no budget change. A possible interpretation is that hardware changes have actual costs associated with the company, therefore, the company  can easily estimate those costs. On the other hand, software cost estimation tend to be zero i.e. no budget changed, which indicates that either the company perceive wrongly Software changes as not costly or it is very difficult to estimate the correct cost (Fig. \ref{fig:Figure_5}).\

\subsection{Software - Hardware intensive requirements VS Analysis Time}
In order to understand the previous finding, a follow up analysis was conducted regarding the analysis time of the Software and the Hardware requirements. The analysis time is the time between the date on which a change request was decided and the last date on which the requirement was changed. Due to the fact that a change in a requirement may occurred before the change request was decided, negative values exist. The time is computed in total days.\

\begin{figure}[ht]
	\centering
	\includegraphics[width= 1\linewidth]{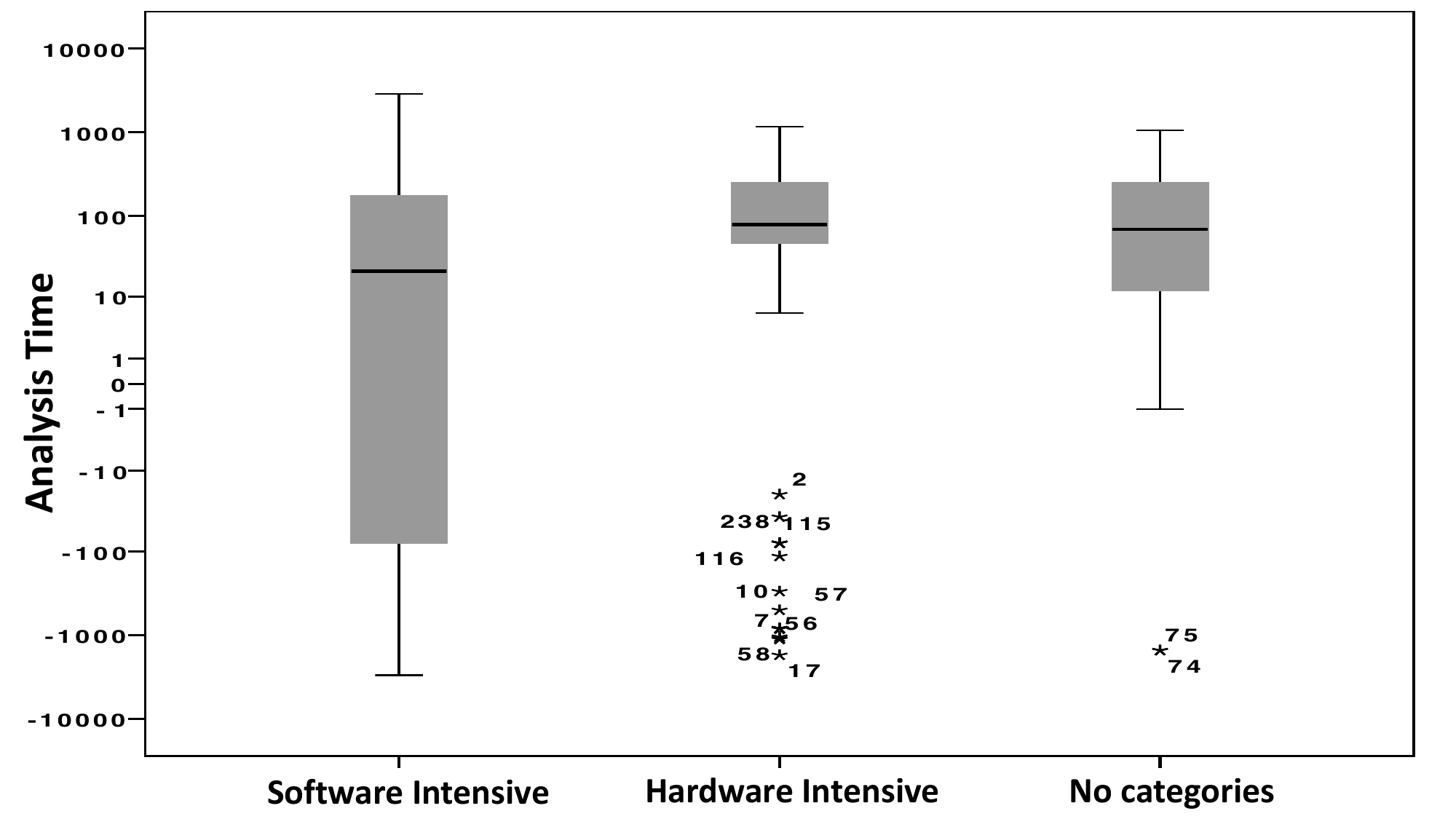}
    \caption{Software - Hardware intensive requirements VS Analysis Time}
    \label{fig:Figure_6}
\end{figure}

The results from the non-parametric test showed that there is statistical significant difference between the analysis time and the software or hardware characteristic of a requirement ($p<0.05$). Moreover, in Fig. \ref{fig:Figure_6} the results showed that analysis time presents more variation for Software rather than for Hardware requirements.  A possible interpretation is that since Software changes are open for longer periods of time, they are harder to handle. Another interpretation could be that the company faces difficulties in managing the software changes and a solution would be to acquirer more input from a outside the company, i.e. a software vendor, for managing software changes satisfactorily.\

\subsection{Number of Comments VS Budget}
\begin{figure}[ht]
	\centering
	\includegraphics[width= 1\linewidth]{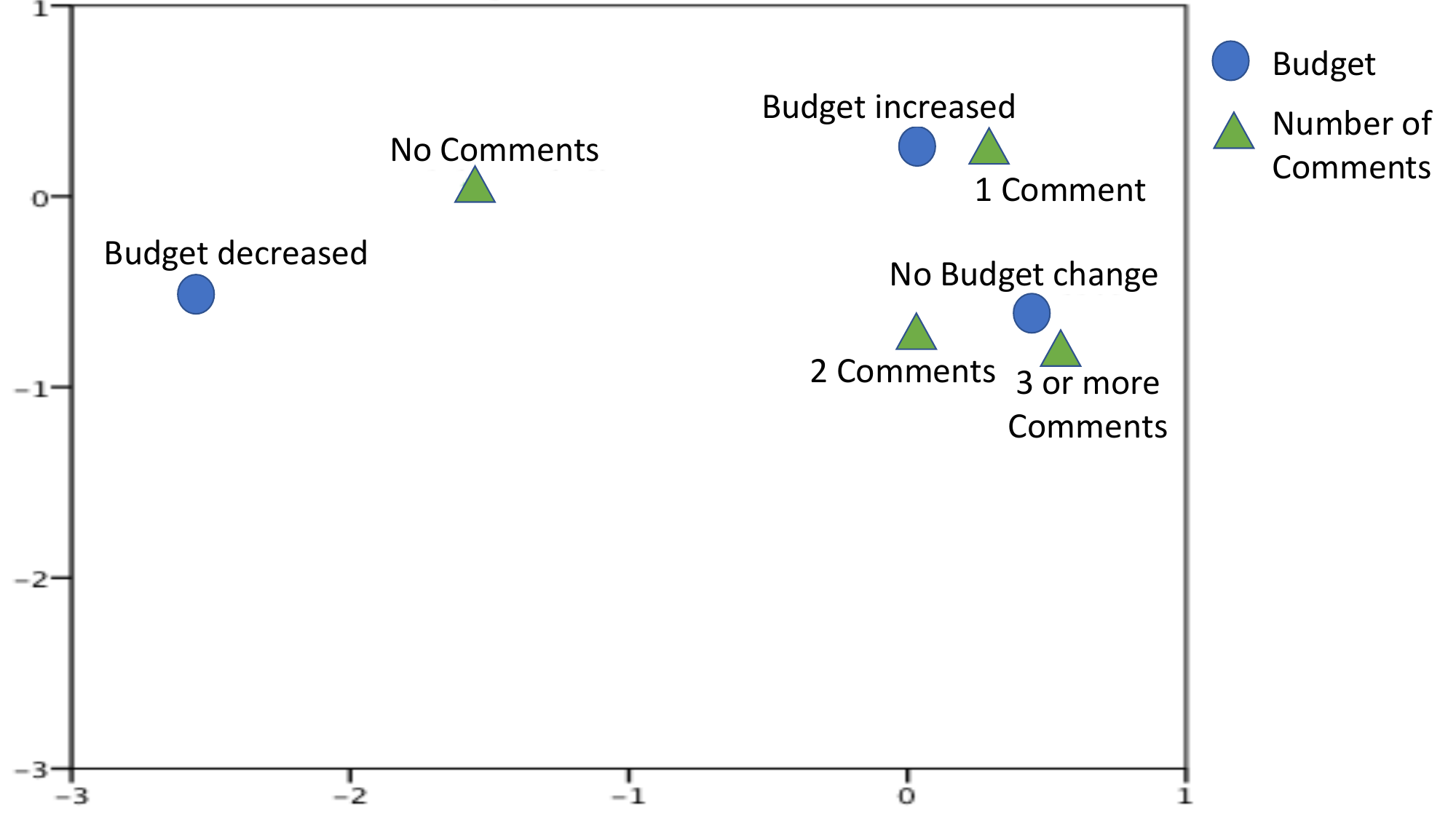}
    \caption{Number of Comments VS Budget}
    \label{fig:Figure_7}
\end{figure}

A chi-square test of independence was conducted between the number of the comments and the budget of the change request. The results showed that Budget is statistical significant associated with the Number of the comments ($p<0.05$, Cramer's $V=0.200$). In particular, the contingency tables show that almost the 80\% of the change requests connected with increased budget, was commented on only once.  On the other hand, among the change requests with no budget change, more than half of them were commented on more than twice. The above results indicated that a long conversation (more comments) on a change request leads us to lower the cost of that change request. The visualization of those results from Corresponding analysis are available at Fig. \ref{fig:Figure_7}.

\subsection{Time sequence of change requests and comments VS Budget}
A follow up analysis was conducted regarding the Time sequence of change requests and comments and the Budget of the change requests. The results from the contingency tables, chi-square test of independence and Corresponding analysis showed that the Budget is associated with the chronological order the comments and the change request happened ($p<0.05$, Cramer's $V=0.300$).  More specifically, in the 80\% of the requirements related with an increased budget, a change request took place before the discussion existed. On the other hand, in almost the 60\% of the requirements related with no budget change, a discussion among the employees existed prior to the change request.

\begin{figure}[ht]
	\centering
	\includegraphics[width= 1\linewidth]{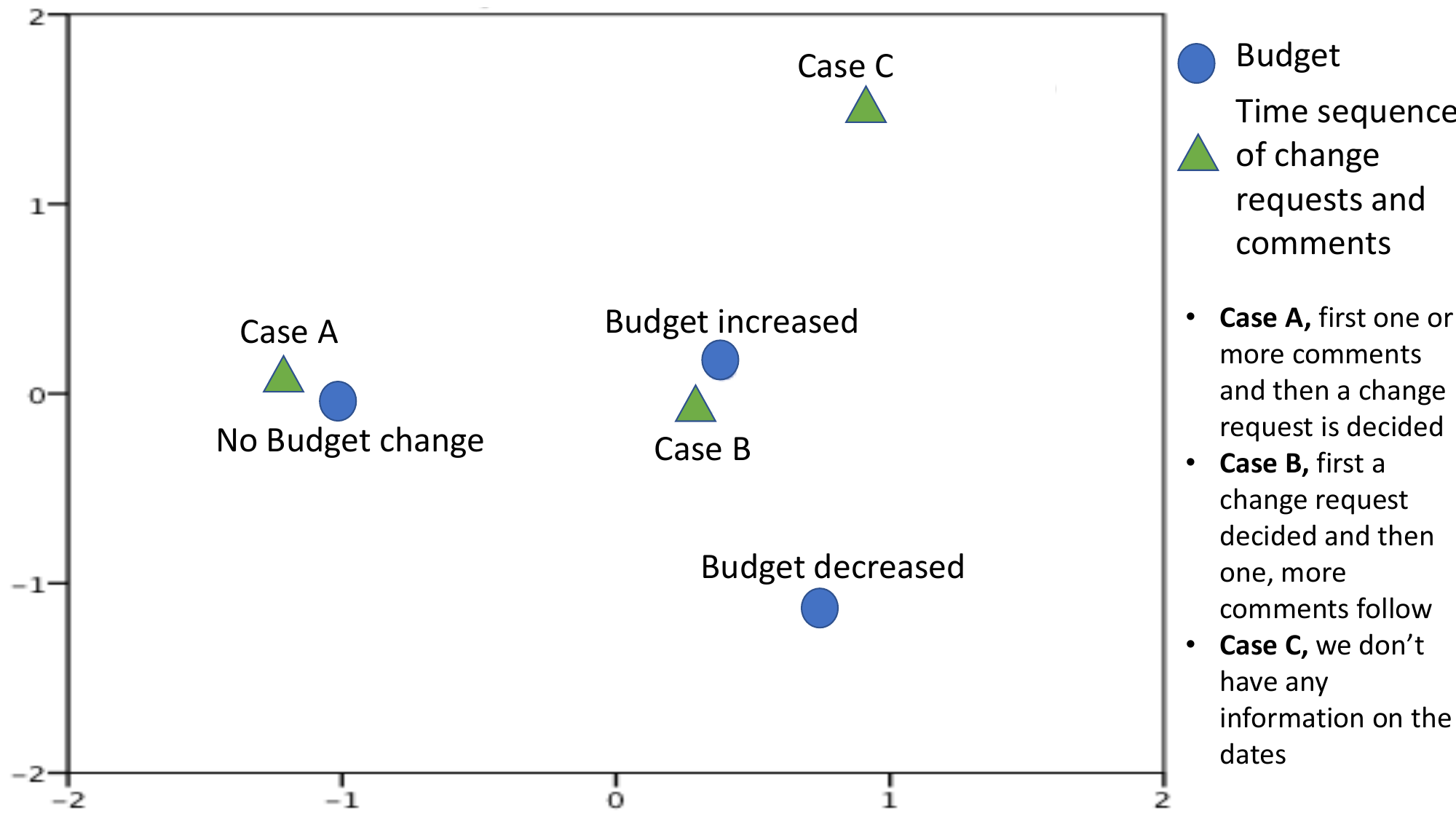}
    \caption{Time sequence of change requests and comments VS Budget}
    \label{fig:Figure_8}
\end{figure}

\section{Discussion-Conclusions}\label{sec:con}
In this case study we investigated the Swedish Transportation Agency (STA), a leading domestic company in the field of transportation and in particular, whether the quality of a requirement has an influence on how it was implemented. The present empirical study is exploratory and focuses on the investigation of whether the requirements' inherent characteristics impact on the project budget and in which ways. Even though the case study is exploratory and the finding cannot be generalized to an isolated company or case study, however they could offer us with a pattern on how the different requirements' characteristics actually impact the project budget.\

The first finding showed that software requirements are associated with less budget when a change request occurs. The results refer only to changes in the budget. Based on the analysis results and the natural language requirements, a possible interpretation is that the estimation of Software budget changes seems to be more difficult in comparison to the Hardware budget changes. Thus, the Software changes requests are "harder" to handle and therefore to be managed since many software requirements budget changes are not even estimated. A potential reason could be that the budget for change requests for Software requirements is not estimated by the employees as analysis costs, probably because they are outsourced to a vendor. \ 

The second follow up finding indicated that the Software changes are open for longer than the Hardware changes. Thus, Software management is more difficult and the company may require input and help from an outside vendor in order to manage Software requirements in a more satisfactory way.\ 

An interesting finding arose regarding employee discussions on the change requests. The more they discuss and interact on a change request, the lower the actual estimated budget change is. A possible interpretation could be that by discussing a change request you may find cheaper solutions for solving the problem i.e potentially find smarter ways with a smaller cost for to the company.\

Finally, when requirements are complemented with comments before requesting a change, the company may "save" money. This may happen due to a better understanding of the needs, i.e. the change requests, instead of immediately asking for a change of the requirements.\ 

The data gathered in studies such as ours are affected by various sources of variation and are therefore subject to large variability. The statistical analysis of such data can reveal significant differences, trends, disagreements and groupings between the practitioners and can constitute a valuable aid for understanding the attitudes and opinions of the stakeholders and therefore a tool for better Software Management.

As a future work, we will continue investigating how do the inherent characteristics of the requirements impact on the project's budget and try to explain the reasons behind this. Our focus is to provide support to companies for improving their ways of managing their Software requirements' changes. In the context of our work, we plan to continue research and efforts towards efficient and effective Software management in Software engineering.

\vspace{25mm}

\section{Acknowledgements}
The work is supported by the KKS foundation through the S.E.R.T. Research Profile project at Blekinge Institute of Technology (BTH). Moreover, it was partially supported by a research grant of the ERSAK project  at Trafikverket, the Swedish Transport Agency (STA). The authors have no conflicts of interests to declare.\

\end{document}